\documentclass[sigconf,authorversion,nonacm,plainpages=false]{acmart}
\usepackage{amsmath}
\usepackage{xcolor}
\usepackage{subcaption}
\usepackage{listings}
\usepackage{framed}
\usepackage{fancyhdr}
\usepackage{tabularx}
\usepackage{graphicx}
\usepackage{booktabs}
\usepackage{enumitem}
\usepackage{tikz}

\definecolor{vgreen}{RGB}{104,180,104}
\definecolor{vblue}{RGB}{49,49,255}
\definecolor{vorange}{RGB}{255,143,102}

\lstdefinestyle{verilog-style}
{
    language=Verilog,
    basicstyle=\small\ttfamily,
    keywordstyle=\color{blue},
    identifierstyle=\color{black},
    commentstyle=\color{vgreen},
    numbers=left,
    basicstyle=\footnotesize,
    numberstyle=\tiny\color{black},
    numbersep=5pt,
    tabsize=2,
    moredelim=*[s][\colorIndex]{[}{]},
    literate=*{:}{:}1
}

\lstdefinestyle{C-style}
{
    language=C,
    basicstyle=\small\ttfamily,
    keywordstyle=\color{black},
    identifierstyle=\color{black},
    commentstyle=\color{black},
    numbers=left,
    numberstyle=\tiny\color{black},
    numbersep=10pt,
    tabsize=8,
    moredelim=*[s][\colorIndex]{[}{]},
    literate=*{:}{:}1
}

\makeatletter
\newcommand*\@lbracket{[}
\newcommand*\@rbracket{]}
\newcommand*\@colon{:}
\newcommand*\colorIndex{%
    \edef\@temp{\the\lst@token}%
    \ifx\@temp\@lbracket \color{black}%
    \else\ifx\@temp\@rbracket \color{black}%
    \else\ifx\@temp\@colon \color{black}%
    \else \color{black}%
    \fi\fi\fi
}
\makeatother

\usepackage{hyperref}
\hypersetup{ colorlinks = true, 
urlcolor = blue, 
linkcolor = blue, 
citecolor = blue 
}

\DeclareUrlCommand\url{\color{blue}}

\begin{document}

\title{Security Properties for Open-Source Hardware Designs}

\author{Jayden Rogers, Niyaz Shakeel, Divya Mankani, Samantha Espinosa, Cade Chabra, Kaki Ryan and Cynthia Sturton}

\begin{abstract}
The hardware security community relies on databases of known vulnerabilities and
open-source designs to develop formal verification methods for identifying
hardware security flaws. While there are plenty of open-source designs and
verification tools, there is a gap in open-source properties addressing these
flaws, making it difficult to reproduce prior work and slowing research. This
paper aims to bridge that gap.

We provide SystemVerilog Assertions for four common designs: OR1200, Hack@DAC
2018's buggy PULPissimo SoC, Hack@DAC 2019's CVA6, and Hack@DAC 2021's buggy
OpenPiton SoCs. The properties are organized by design and tagged with details
about the security flaws and the implicated CWE. To encourage more property
reporting, we describe the methodology we
use when crafting properties.

\end{abstract}


\maketitle

\section{Introduction}

``For better or worse, benchmarks shape a field''--David Patterson~\cite{patterson2012better}.

The use of formal verification methods for the security validation of hardware
designs is an active area of research. The research has led to the development
of new tools~\cite{ryan2023sylvia,meng2021rtl,anton2023fault,muller2021formal}
and uncovered new security bugs and
vulnerabilities~\cite{zhang2018end,meng2021rtl,anton2023fault}. In support of
these efforts, the hardware security community has developed and
published a large number of open-source hardware designs for tool development (e.g.,~\cite{or1200,schiavone2018quentin,zaruba2019cost}), along with
flawed designs for use as benchmarks in the evaluation of any new methodology
(e.g.,~\cite{hackatdac,trusthub}).

\begin{figure}[hbt!]
    \centering
    \includegraphics[width=0.45\textwidth]{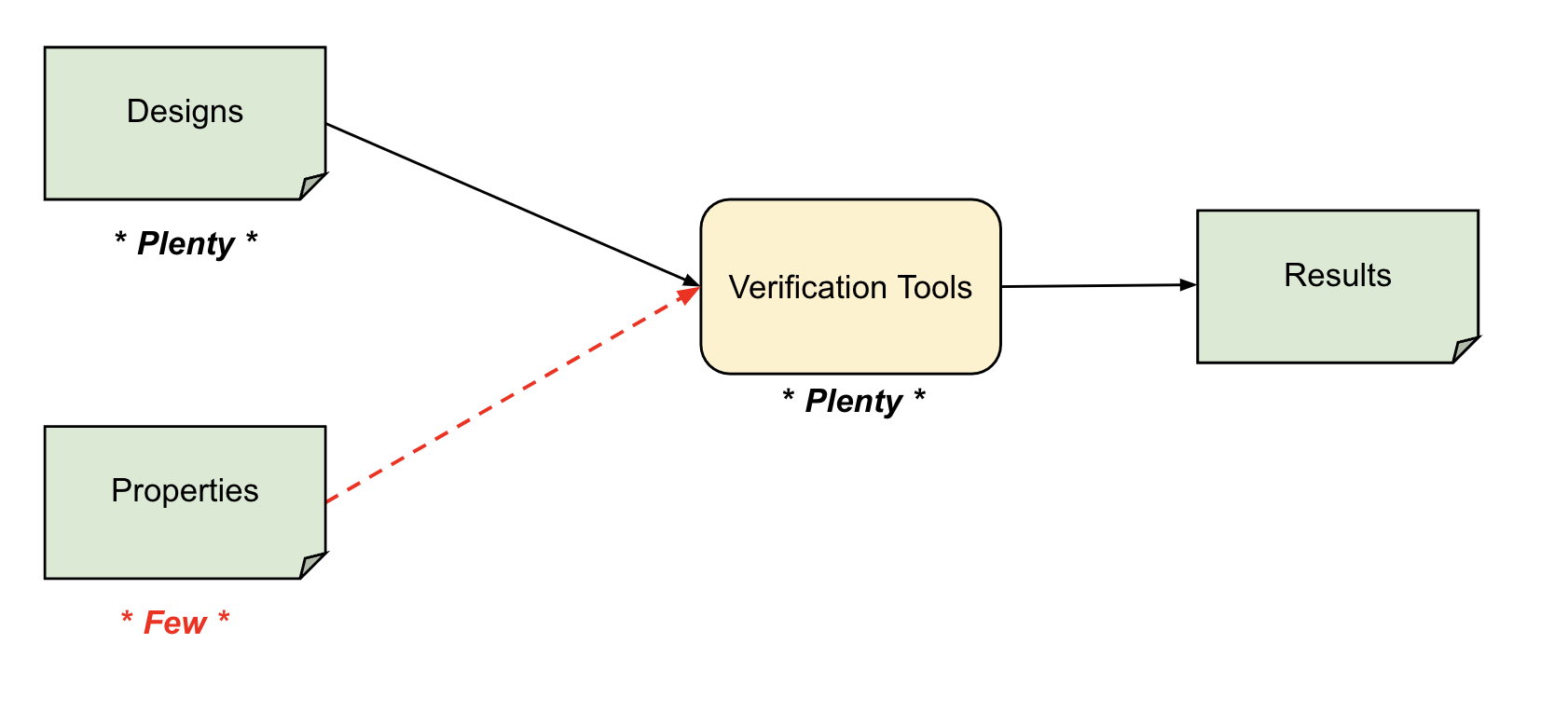} 
    \caption{In the hardware verification community there is no shortage of open-source designs or verification tools,
    but the verification properties still remain scarce, hard to find, and hard to write. This work aims to bridge that gap
    through both the contribution of an open-source database of properties, a methodology for property writing
     , and a call for authors to start providing the properties they use in their work.}
    \label{fig:property-gap}
\end{figure}

Unfortunately, the security properties needed for verification are not similarly
available. Many of the new methods being developed are property-based, where
the verification engine takes a (possibly buggy) hardware design and a
set of properties as input and outputs any found counterexamples (CEX) to the
properties. There is general consensus that developing properties is a
laborious task~\cite{tyagi2022thehuzz,kastner2022automating},
with ongoing research into generating properties
automatically~\cite{zhang2017identifying,deutschbein2020evaluating,zhang2020transys},
most recently using large language
models~\cite{paria2023divas,dipu2023agile,kande2023llm}. Still, there remains a
lack of well-vetted, publicly available, formally written properties tied to a
particular snapshot of a design (see Figure \ref{fig:property-gap}). This work is a step toward providing that
resource.

Our contribution is largely pragmatic. 
We write four sets of security properties for four buggy designs
commonly used in the literature: the OR1200 OpenRISC CPU core~\cite{or1200}, the
PULPissimo SoC with RISC-V core~\cite{schiavone2018quentin}, and two versions of the OpenPiton
SoC with the CVA6 RISC-V core~\cite{zaruba2019cost,balkind2016openpiton}.

The properties are written to find known security bugs in each design. For
OR1200, we create a snapshot of the design with a collection of known bugs taken
from the literature inserted~\cite{zhang2018end,hicks2015specs,ryan2023sylvia};
for the PULPissimo and OpenPiton SoCs, we use the buggy designs used in the
Hack@DAC competitions in 2018, 2019, and
2021~\cite{pulpissimo2018hack,cva62019hack,cva62019hackbugs,openpiton2021hack}. The
properties are written as SystemVerilog Assertions (SVA), the industry standard
for trace property specification~\cite{systemverilogstandard}, and the security
bugs they cover are categorized according to the CWE taxonomy~\cite{cwe}. We
provide this categorization for OR1200 and supplement the existing
categorization for the Hack@DAC designs.

We write 71 properties for the
OR1200, successfully detecting all known bugs and offering additional security
assurance based on requirements from prior work. For PULPissimo, we create 20
properties to address 31 known bugs, an improvement over prior efforts that
reported (but did not publish) 15 properties. Finally, we design 11 and 20 properties for the 66 and 99 known bugs in the 2019 and 2021 OpenPiton SoC designs, respectively.
We use JasperGold, the industry-standard formal verification tool from Cadence,
to confirm that the properties find the known bugs. The properties, along with a snapshotted version of the buggy open-source design they target, are available as benchmarks for future evaluation of verification tools.

To further motivate the need for open-source properties, we present a case study
illustrating the challenges to reproducibility that arise when properties are
missing from the public record. To encourage more property reporting, we
describe the methodology we use to craft properties so that others can follow
the same steps. Our hope is that together the properties and methodology will accelerate and
improve future evaluation of formal verification tools.

This paper makes three contributions.
\begin{itemize}[noitemsep]
\item Provide well-vetted properties written for four buggy designs
  commonly used in the
  literature along with their associated snapshotted designs:
  \url{https://github.com/HWSec-UNC/verification-benchmarks}
\item Demonstrate the importance of properties for reproducibility through a
  case study of the community-standard Hack@DAC 2018 buggy PULPissimo design.
\item Describe the methodology we use to craft properties to encourage others to
  contribute new properties. 
\end{itemize}

\section{Background}


We discuss related work and provide context for our work.

\subsection{Scope}
Our focus is on research using formal verification methods for the security
validation of hardware designs; functional validation is out of scope. The line
between security properties and functional properties can be blurry, as security
flaws often stem from functional bugs in the design.  However, the resources we
consider are designed specifically for security validation activities aimed at
identifying vulnerabilities exploitable by an attacker. These resources may or
may not be appropriate for functional validation activities.

\subsection{Open-Source Resources}
We are contributing formally stated security
properties used during the formal verification process. Here, we discuss
the resources currently available.

\emph{Trust-Hub}.
TrustHub is an NSF-funded website designed to support research and
education in hardware security~\cite{trusthub}. The website serves as a directory for various hardware security resources and also offers resources directly, including Trojan-infected design components, taxonomies, and datasets related to physical hardware security. Two resources found on the Trust-Hub site are the
Security Property\slash Rule Database and the SOC Vulnerability Database. We
discuss each here.

The Security Property\slash Rule
Database~\cite{farzana2019soc} contains Verilog and
SystemVerilog modules implemented at the register transfer level (RTL) or gate level, along with
a set of SVA properties for verifying each module. The database is useful,
providing both the design and formally specified properties for each
design~\cite{farzana2019soc,farzana2021soc}. However, this database is only a
first step. At the time of this writing, the database includes a handful of
properties for components of the CV32E40P RISC-V core. The rest are
individual design components, such as a write-once register or FIFO block or
smaller modules such as an SPI controller, RSA module, or AES module. Each design is accompanied by one to five properties.

The SoC Vulnerability Database~\cite{socvulndb} lists six vulnerabilities for
modules in the CVA6 RISC-V CPU. The vulnerability is associated with the appropriate CWE
entry, and for each vulnerability, one to four SVA properties are
provided. However, the particular version of the CVA6 design for which the properties were
written is not mentioned, and the properties may require significant changes to
be useful for the current version of the design. 

Our work expands on these two databases at a
larger scale, providing 71, 20, 11 and 20 properties, respectively, for
four different buggy CPU and SoC designs, along with snapshotted versions of the
designs for which the properties were written. (After publication we will submit our repository for
inclusion in the Trust-Hub directory to make it easy for the community to find
the resource.)

\emph{Hack@DAC Competitions}.
The Hack@DAC competitions \cite{sadeghi2021organizing,hackatdac} have been a
boon to the hardware security community. In the annual competition, competitors
are given the RTL of an SoC design and must find the inserted security bugs. The
organizers are a team of academic researchers and industry practitioners, and the
bugs inserted reflect real-world vulnerabilities.

The design used in the 2018 instance of the competition is a PULPissimo
SoC~\cite{pulpissimo2018hack} and has become a standard benchmark for hardware
security research (e.g.,~\cite{ryan2023sylvia,kande2023llm,muller2021formal}), thanks in part to the
organizers providing natural-language
descriptions of the bugs inserted into the design~\cite{dessouky2019fails}.

Recently, the OpenPiton SoC designs used in the 2019 and 2021 competitions were made available
as well~\cite{cva62019hack,openpiton2021hack}, along with the bug descriptions
for the 2019 version~\cite{cva62019hackbugs} and an overview of the 2021 version~\cite{chen2022trusting}.

However, the available resources stop short of pinpointing the actual bugs in the
RTL. Research teams using these designs as a benchmark for the development or
evaluation of their tool must first establish ground truth by finding the bug in
the RTL through whatever means possible. The SVA properties we provide will
boost the usefulness of these benchmark designs by pinpointing the bug in
the design and acting as a baseline set of assertions for use in future research.

\emph{CWE Database}.
The Common Weakness Enumeration (CWE) database from MITRE is a
community-maintained resource that categorizes common flaws in
software and hardware that often lead to vulnerabilities~\cite{cwe}. Entries in the
database describe the flaw, provide information about how a vulnerability
may manifest, and may include example code snippets. The database can be used
as a guide when writing the formally stated properties needed for security
verification~\cite{gogri2021texas,restuccia2021aker,cycuitycweguide}. We used
CWEs as a guide when writing assertions and link to the appropriate CWE for the
inserted bugs when they were missing.

\subsection{Related Work}
We first discuss work using formal verification for the security validation of
hardware designs with a focus on how and whether SVA properties are used and
made available. We then briefly discuss work using simulation-based security
validation tools that may or may not use SVA properties.

\emph{Property Generation}.
A variety of techniques have been used to automatically generate security
properties for hardware designs. The most recent papers explore using large language model (LLM)
code generation to generate RTL
properties~\cite{paria2023divas,kande2023llm,meng2023unlocking,assertllm2024,llm4dv2023,assertllm2024}. Other
techniques include cross-design property translation, mining properties from
design behavior, and using CWE descriptions to generate
properties~\cite{zhang2017identifying,zhang2020transys,dipu2023agile,deutschbein2021isadora,deutschbein2020evaluating,deutschbein2021mining,htpgfv2024}.

While some papers provide a handful of the properties
generated~\cite{paria2023divas,deutschbein2021isadora,dipu2023agile,deutschbein2020evaluating,kande2023llm,deutschbein2021mining},
we found only one
paper that provides the complete set of generated
properties~\cite{zhang2020transys}. Furthermore, the automated approaches
typically generate thousands of properties with varying degrees of
relevance. Selecting the most relevant properties is done either manually or
through statistical analysis. Both approaches benefit from having a set of
known-good properties as a starting point.

\emph{Property-Based Formal Verification}.
Many new analyses and frameworks
for formally verifying security properties of a design have been
developed~\cite{meng2021rtl,zhang2018end,hicks2015specs,ryan2023sylvia,ryan2023sylvia,meng2023ensuring,fang2023wasim,7927266,lyu2020automated}. However,
many papers do not report the properties used in the evaluation. A few
do~\cite{hicks2015specs,solem2023applying,8587741}, and a few more report one or two of the properties
used~\cite{anton2023fault,rajendran2023hunter,rajendran2024exploring}.
But if we consider only
papers using the industry-standard SVA for property specification, we are left
with few
properties~\cite{rajendran2023hunter,rajendran2024exploring,solem2023applying}, making comparisons between tools difficult. The
properties we provide will make it easier for new research in security verification techniques to make comparisons
with the existing state of the art.

\emph{Fuzzing}.
In recent years, fuzzing has become a popular research focus within the hardware security community~\cite{trippel2022fuzzing,canakci2021directfuzz,tyagi2022thehuzz,chen2023hypfuzz,xu2023morfuzz,chen2023psofuzz,canakci2023processorfuzz,gohil2023mabfuzz,sugiyama2023surgefuzz,hossain2023socfuzzer,hossain2023taintfuzzer}. Fuzzing is a simulation-based method, but unlike testing, fuzzing benchmarks do not include expected output values. Rather, to find bugs, fuzzing tools use a golden reference
model, security properties, or some combination of the two. The SVA security
properties we provide in Section~\ref{sec:properties} may be suitable for use with some fuzzers. However,
in our examination of the literature, we observed that many new hardware fuzzing
tools use the golden reference model technique. While this 
technique has its own limitations, it does avoid the challenges and drawbacks of property writing 
discussed in this paper.

\emph{Taint Tracking}.
Taint tracking is another simulation-based method that has seen both academic~\cite{solt2022cellift,meza2023security}
and commercial~\cite{cycuity} success in the hardware security community. (See Hu
et al. for a survey~\cite{hu2021hardware}.) These tools
work by first instrumenting the design with tracking logic and then analyzing how information flows through the design
during simulation. Taint tracking tools
typically have their own specification language for writing information-flow
properties; the SVA properties we provide here would not be suitable for use
with those tools. Although not a formal verification method, taint tracking also requires well written
properties, and there have been efforts toward easing the property writing process~\cite{restuccia2021aker,deutschbein2021isadora}.


\section{Property Development}
\label{sec:properties}
We develop properties for four processor and SoC designs. 
The properties are all trace properties specified in the SystemVerilog Assertion language (SVA). 
Additionally, the
properties are all safety properties, meaning the properties are
violated when an undesirable state or finite sequence of events is
reached.

We provide the complete set of properties and design information in a public repository: ~\url{https://github.com/HWSec-UNC/verification-benchmarks}. The
properties are organized by the design they were written for, and the repository is set up to allow pull
requests to allow others to contribute or make corrections.

In the following sections, we document our methodology for 
manually generating these properties, discuss the hardware designs studied, and
describe a few illustrative properties for each.

\subsection{Methodology}
\label{sec:methodology}
When developing a property, we 
are operating in one of two settings: 
\begin{enumerate}
	\item The goal is to write a new property or set of properties capturing desired secure behavior of a design.
	\item The goal is to write a new property or set of properties capturing specific known buggy behavior of a design. 
\end{enumerate}
In setting \#1, we have a design and the objective is to capture and specify, in the form of SVA
assertions, the desired secure behavior the design should provide. When we
verify properties of this type for the design, there is no expectation of finding an assertion violation. A secure and
bug-free design, for example, would see no violations of such properties. For
the OR1200, we develop properties according the OpenRISC 1000 specification
document. Examples of this are discussed in \ref{sec:or1200}.

In setting \#2, we have a design with a known vulnerability or bug and the
objective is to write
SVA assertion(s) to catch the bug. When the buggy design is verified, we expect
to find a violation of our property. For the Hack@DAC 2018, 2019 and 2021
designs, we take this approach; see sections \ref{sec:pulpissimo-properties} and
\ref{sec:hack-other} for more details.

Ideally, a complete specification developed in setting \#1 would be
sufficient to capture all bugs. In reality it is helpful to work in
both settings to build a more complete set of properties.

Our process for writing properties is as follows:
\begin{enumerate}
\item \textbf{Study Design Specifications and Survey Literature}: 
The first step in the process is to study the architecture of the design of interest, 
paying close attention to security-specific elements like privilege (de)escalation
mechanisms and instructions or flags that modify control flow. 
If available, we review and study previously established properties 
related to the particular design or similar designs. Some works that we commonly refer back to in our property development
process include \cite{meng2021rtl}, \cite{dessouky2019fails} and \cite{tarek2023benchmarking}. These sources 
offer insights into recurring issues or potential vulnerabilities in the designs
that would be good to target with a formal property. 

\item \textbf{Investigate Bug Descriptions}: If starting with a bug description
  (setting \#2), the next step is to identify the modules and signals relevant to the bug. 
One challenge is that bug descriptions are in high-level English and finding direct mappings to RTL code
is not always straighforward. Additionally, two property writers may produce two
different properties given the same bug description. 
Property writing should be treated as an iterative process where multiple properties for the same bug can be compared, combined, and 
refined to capture the buggy behavior as comprehensively as possible. 

\item \textbf{Research Matching CWEs and RTL Examples}: For known bug
  descriptions (setting \#2), we look for corresponding Common Weakness
  Enumerations (CWEs). This mapping can provide a clearer understanding of the
  vulnerability and guide the investigation.
If a CWE match is identified, we search for example source code from similar
designs that might offer hints. These examples
can shed light on potential approaches to writing an assertion.

\item \textbf{Write the Property}: At this point we write a first draft of the
  property using the information we've gathered. We specify the expected behavior we are trying
  to capture with our properties in the form of
  SystemVerilog assertions. These assertions are written in a combination of boolean and temporal logic. Each 
  assertion is associated with a particular clock domain, and values of signals are sampled at clock edges 
  (ex. \texttt{@(posedge clk)}) to ensure synchronization with the design's timing.
  The process is iterative, and we often come back to the writing step after our first attempt to verify the property.
\item \textbf{Verify the Property}: The final step is to attempt to verify the
  property. When in setting \#2 we expect to find a violation. If none is found,
  we iterate by studying the design more deeply and editing our property. When
  in setting \#1, the absence of a violation does not necessarily indicate the property is sound. Determining when to stop and feel confident in the property’s validity requires judgment, multiple sets of eyes and practice.
\end{enumerate}

In \ref{sec:examples} we will highlight our experiences writing properties using our methodology, along with some of the challenges and 
obstacles we faced.


\subsection{Designs Studied}

One challenge we encountered during our development is that a property written for one version of a design may not be
applicable to newer versions of the design. Clock-cycle timing, signal naming,
and data transfer between registers may all be affected by edits to a design
even if the high-level behavior of the design related to a property remains unchanged. To address this, 
we provide documentation in our benchmarks repository for each design that we write properties 
for. We specify the version, commit point of the design in its home repository, and a static snapshot of the design. 
\subsubsection{OR1200}
\label{sec:or1200}
The OR1200 is a 5-stage pipelined single-core processor~\cite{or1200} with a
relatively long history of use in the literature for evaluation of formal verification techniques~\cite{hicks2015specs,zhang2018end,ryan2023sylvia}. Security and functional bugs found in the design over time have been documented on the OpenCores~\cite{opencores}, Bugzilla~\cite{bugzilla} and Mitre CWE databases~\cite{cwe},
 as well as in issues opened by developers on their GitHub repository~\cite{or1200}. Additionally, there are cases where security researchers have identified native bugs in one generation of the processor that a subsequent version or commit 
claims to have resolved~\cite{zhang2017identifying}, but the bugs reappear or persist over time~\cite{zhang2018end}.
 
 We have created a benchmark that snapshots the OR1200 in a buggy state
 containing 31 security bugs collected from prior
 work~\cite{hicks2015specs,zhang2017identifying}. 
Along with the design, we include a set of 71 security properties that target behaviors related
to the inserted bugs. Fifteen of these SVA properties are inspired by OVL
properties provided in prior work~\cite{hicks2015specs}; the remaining
properties are based on properties described in prior
work~\cite{bilzor2011security,zhang2017identifying,zhang2020transys}.  
Despite prior use of OR1200, the community still lacks an available set of SVA properties
written for the design and a buggy benchmark that can
serve as ground truth during evaluation. This
benchmark fills that gap.


\subsubsection{Hack@DAC 2018}
\label{sec:pulpissimo-properties}

The PULPissimo SoC is a RISC-V-based SoC with a 6-stage pipeline and several security enhancements.
The PULPissimo is the design that is used in the
HACK@DAC 2018 competition~\cite{pulpissimo2018hack} and is the basis for the
HardFails paper~\cite{dessouky2019fails}.  
The version of the design used in the HACK@DAC 2018 competition has a set of
bugs associated with it, some native to the design and others inserted by the competition organizers.

We developed a set of properties targeting the bugs known to be
present in the version used in the Hack@DAC competition. In writing these properties, we used the bug descriptions available in the
literature~\cite{pulpissimo2018hack,dessouky2019fails,meng2021rtl}, and map each bug to the appropriate CWE.
In our repository, we provide a snapshot of the HACK@DAC 2018 design that we used to develop these
properties, the full set of SVA properties, and an associated CWE for each property.


\subsubsection{Hack@DAC 2019 and 2021}
\label{sec:hack-other}

CVA6 (formerly known as the Ariane SoC) is another open-source SoC design that has grown
 increasingly popular within the systems and security research communities. 
The designs used for the HACK@DAC 2019 \cite{cva62019hack}
and 2021 \cite{cva62021hack} competitions were based on the CVA6 architecture. 
These two designs have 66 and 99 bugs inserted, respectively. Some of these 
vulnerabilities were native to the design and some, similar to HACK@DAC 2018, were inserted
by the competition organizers. 

Using bug descriptions and case studies published in the literature, we develop our
own properties for finding the bugs. 
We snapshot each design and include the version used for verification with JasperGold in the 
benchmarks repo. These snapshots are slightly modified from the versions in the official 
HACK@DAC repositories. To successfully check our assertions in JasperGold, we had to resolve
a series of syntax errors and make small fixes to compile the entire SoC. 
We include these finalized versions in our benchmarks repo along with 11 and 20 properties
for each design, respectively. 

\subsection{Property Examples}
\label{sec:examples}
To demonstrate our methodology, as well as the general challenges associated with
property writing, we provide some examples of properties we developed as
part of the set of benchmarks. 

First, given only a bug description, it is challenging to pinpoint the bug in the code
without deep knowledge of the code base comparable to what the original
design engineers would have. Whenever possible, we looked for agreement
with our findings in the literature. For example, Bug 76 from Hack@DAC
2021 is described as ``Some of the register lock registers are not locked by
register locks''~\cite{openpiton2021hack}. Our investigation led us to the code
shown in Listing~\ref{lst:bug76hackdac21}, with the apparent bug on Line
2. We found confirmation of this bug from a description of a similar bug by
Ahmad et al.~in their work on CWE Analysis~\cite{ahmad2022dont}.

\begin{lstlisting}[xleftmargin=.01\textwidth, %use left and right margins to center listing
    xrightmargin=.005\textwidth, %use left and right margins to center listing
    caption={The buggy line of code \texttt{reglk\_wrapper} module of Hack@DAC
      2021 OpenPiton SoC},
    label={lst:bug76hackdac21},
    frame=single,
    float,
    style={verilog-style}]
else if(en && we)
 case(address[7:3])
  0: reglk_mem[0] <= reglk_ctrl[3] ? reglk_mem[0]:wdata;
  1: reglk_mem[1] <= reglk_ctrl[1] ? reglk_mem[1]:wdata;
  2: reglk_mem[2] <= reglk_ctrl[1] ? reglk_mem[3]:wdata;
  3: reglk_mem[3] <= reglk_ctrl[1] ? reglk_mem[3]:wdata;
  4: reglk_mem[4] <= reglk_ctrl[1] ? reglk_mem[4]:wdata;
  5: reglk_mem[5] <= reglk_ctrl[1] ? reglk_mem[5]:wdata;
  default: ;
 endcase
\end{lstlisting}   

Similarly, many of the bugs hinged on deep knowledge of the 
peripherals and how they interfaced with the memory interfaces in the SoC. It seemed 
to be the case that each peripheral was assigned a certain index in memory for  
configuration purposes. There was a bug in 2019 that described: ``Processor access to CLINT 
(core level interrupt controller) grants it access to PLIC (processor level interrupt controller)
regardless of PLIC configuration.'' This ultimately came down to understanding the hardcoded
indices in the RTL in Listing~\ref{lst:bug1hackdac19}. PLIC corresponds to 6 and CLINT corresponds to 7. 

\begin{lstlisting}[float,xleftmargin=.01\textwidth, %use left and right margins to center listing
    xrightmargin=.005\textwidth, %use left and right margins to center listing
    caption={The buggy line of code \texttt{axi\_node\_intf\_wrap} module of Hack@DAC
      2019 OpenPiton SoC},
    label={lst:bug1hackdac19},
    frame=single,
    style={verilog-style}]
for (i=0; i<NB_SUBORDINATE; i++)
 begin
 for (j=0; j<NB_MANAGER; j++)
  begin
   assign connectivity_map_o[i][j] =  
           access_ctrl_i[i][j][priv_lvl_i] || 
           ((j==6) && access_ctrl_i[i][7][priv_lvl_i]);
  end
 end
\end{lstlisting}

This same bug is present in the 2021 design, but we were unsuccessful in writing a property for it 
at the time of submission. The source code was updated, and we couldn't locate the corresponding logic in the RTL,
even though we were able to enumerate the peripherals in the new generation of the CVA6 SoC.
This challenge arose more than once: a bug description that was the same
or similar between the 2019 and 2021 versions of the same design did not necessarily correspond to property reuse.
The signal names, related modules, and RTL often varied considerably between the two generations, 
making 1:1 mappings difficult. 

Many bug descriptions directed us to a module or area of the 
design, but we then had to decode arbitrarily complex FSM states and transitions. For example,
one bug in the 2021 design has the description: ``Able to write using JTAG without password.'' 
We know we care about the JTAG module and the password, but no more.
Once we locate these assets, we must dig into the FSM of the design to
understand the behavior. The process is slow and tedious, taking many hours for
a large design.

A snippet of the FSM for the JTAG unit is shown in Listing~\ref{lst:bug2hackdac21}. For this bug in particular,
the problem is that we are able to go from IDLE to WRITE without checking that the pass\_check == 1.
There are many cases where being able to write an expressive property is predicated on understanding the FSMs
of the design. Furthermore, FSMs are implementation-specific, and having to target behavior captured by an FSM
 vs. implementation-agnostic security specifications can be challenging. This again demonstrates why
properties are closely coupled to the design they 
were originally written for.  

While bug descriptions may sometimes lead to the correct module, they may also erroneously lead to 
a module or set of signals that end up being red herrings. It is tempting to search based on keywords in the bug description, 
but one really needs a deep understanding of what is 
 going on functionally in the RTL and have some view into what the drivers are 
 for the security-critical signals. Commercial tools like JasperGold offer waveform viewers and 
 menus that facilitate this analysis within a module with relative ease. However, such analysis can
  become complex when spanning multiple modules. Obtaining a comprehensive understanding 
  of data flow and signal drivers can be challenging without a commercial-grade simulation environment.

\begin{lstlisting}[float,xleftmargin=.01\textwidth, %use left and right margins to center listing
    xrightmargin=.005\textwidth, %use left and right margins to center listing
    caption={Buggy FSM in  JTAG module of 2021 OpenPiton SoC},
    label={lst:bug2hackdac21},
    frame=single,
    style={verilog-style}]
case (state_q)
 Idle: begin
  // make sure that no error is sticky
 	if(dmi_access && update_dr &&
      (error_q == DMINoError)) begin
   // save address and value
   address_d = dmi.address;
   data_d = dmi.data;
   if((dm::dtm_op_e'(dmi.op) == dm::DTM_READ) &&
       (pass_check | ~we_flag == 1))
    begin
     state_d = Read;
    end
    else if((dm::dtm_op_e'(dmi.op) == dm::DTM_WRITE) &&
             (pass_check == 1))
    begin
     state_d = Write;
    end
    else if(dm::dtm_op_e'(dmi.op) == dm::DTM_PASS)
    begin
     state_d = Write;
     pass_mode = 1'b1;
    end
    // else this is a nop and we can stay here
  end
 end
\end{lstlisting}

Bug descriptions may not always provide clarity about the level of abstraction where the vulnerability exists. In SoC repositories, 
there is often low-level system/C code that is included where bugs appear to be beyond our scope or involve software-related issues.
Some of the CWEs associated with the HACK@DAC '19 and '21 designs have been tagged as outside the scope of RTL analysis\cite{ahmad2022dont}, 
but relying solely on the bug description, this is not immediately clear. For example, bug 4 (CWE 1191) and
 bug 25 (CWE 1268) both require functional simulation to reproduce. In other words, the vulnerabilities are only 
 realizable at runtime, making property writing more challenging.

\section{Case Study: Examining the role properties play in reproducibility}



The buggy PULPissimo SoC used in the first year of the Hack@DAC competition in 2018~\cite{pulpissimo2018hack}
has become a standard benchmark for hardware
verification ~\cite{meng2021rtl,muller2021formal,ryan2023sylvia,ryan2023sylvia,solem2023applying,kande2023llm}.
This is largely due to the 2019 HardFails paper which details
the design, its security bugs, and provides a link to the public repository hosting the buggy design snapshot ~\cite{dessouky2019fails}.

However, to our knowledge, no publicly available dataset of properties exists for detecting bugs in the design.
Some works include one or two properties~\cite{muller2021formal, paria2023divas},
while others offer only natural-language description ~\cite{meng2021rtl}. 
There is no expectation in the community for verification tool developers to provide 
SystemVerilog Assertions or equivalent RTL-level properties for identifying security flaws.

As a result, every research group that would like to use the 2018 Hack@DAC buggy
Pulpissimo design in either the development or evaluation of their research,
must first develop the particular properties they will need to find the bugs in
the design. This is not a straightforward process and deep expertise of the design. 
Writing a strong property requires knowing first, which modules of the design to target, which signals
within those modules to target, and what the correct behavior of those signals within
each clock cycle. Furthermore, two different research groups are likely to
develop two different sets of properties to target the same set of bug
descriptions. This makes any comparison between reported evaluation results
challenging.  

We can use the Hack@DAC 2018 PULPissimo design to demonstrate the challenge of reproducing
verification results when the properties are not available. For every bug in the
PULPissimo SoC used in the 2018 Hack@DAC competition and described by the HardFails paper~\cite{dessouky2019fails}, we wrote or attempted
to write an SVA property that will find the bug. We then used the Cadence
JasperGold verification engine to look for assertion violations
for the buggy design. Specifically in this case study, we use JasperGold Formal 
Path Verification or \textit{FPV}.  The HardFails authors performed an evaluation using FPV, and used the
 same version of the PULPissimo SoC~\cite{pulpissimo2018hack}. Two examples of assertions are shown in
Listings ~\ref{lst:bug3hack18} and ~\ref{lst:bug13hack18}. 

\begin{lstlisting}[xleftmargin=.01\textwidth, %use left and right margins to center listing
    xrightmargin=.005\textwidth, %use left and right margins to center listing
    caption={Assertion for Bug 3 (Processor assigns privilege level of execution incorrectly from CSR.) in HACK@DAC 2018},
    label={lst:bug3hack18},
    frame=single,
    breaklines=true,
    style={verilog-style}]
assert(~((riscv_core.cs_registers_i.priv_lvl_n == riscv_core.cs_registers_i.PRIV_LVL_M) && riscv_core.cs_registers_i.mstatus_n.mpp == riscv_core.cs_registers_i.PRIV_LVL_U)))
\end{lstlisting}

\begin{lstlisting}[xleftmargin=.01\textwidth, %use left and right margins to center listing
    xrightmargin=.005\textwidth, %use left and right margins to center listing
    caption={Assertion for Bug 13 (Faulty decoder state machine logic in RISC-V core results in a hang.) in HACK@DAC 2018},
    label={lst:bug13hack18},
    frame=single,
    breaklines=true,
    style={verilog-style}]
assert(riscv_controller.id_stage_i.controller_i.ctrl_fsm_ns == riscv_controller.id_stage_i.controller_i.DECODE) |=> (riscv_controller.id_stage_i.controller_i.ctrl_fsm_ns != riscv_controller.id_stage_i.controller_i.DECODE))
\end{lstlisting}

Table~\ref{tab:reproduce-hardfails} shows the results of our reproducibility study. We group each bug by the CWE. We report whether we were able
successfully find the known bugs with our hand-written assertions using JasperGold FPV.
We compare our success in finding bugs to the corresponding JasperGold FPV results in the HardFails paper.

Despite using the same tools and the same design, we report finding bugs that they report
finding only through manual inspection or not finding at all. The reverse is
also true; the HardFails paper reports finding bugs using JasperGold that we
were unable to find or write a property for. Given that the design and verification engine are the same,
we conclude that this difference in outcomes likely results from using a different set of
SVA properties, and points to the importance of making properties open-source
for reproducibility.

\begin{table}[htp!]
  \centering
\begin{tabular}{c c c c}
    \toprule
    CWE ID & \multicolumn{2}{c}{No. of Bugs Found} & Total Bugs Present \\
        \cmidrule(lr){2-3}
           & Our Results & HardFails Results & \\
    \midrule
        1203 & 3 & 3 & 3 \\
        1207 & 2 & 2 & 2 \\
        1206 & 1 & 1 & 1  \\
        1257 & 0 & 1 & 1 \\
	    20 & 1 & 1 & 1\\
	    1221 & 1 & 1 & 1 \\
         1298 & 1 & 1 & 1 \\
         1329 & 2 & 1 & 3\\
	    1245 & 3 & 3 & 3\\
	    1247 & 1 & 1 & 1\\ 
	    1419  & 1 & 0 & 1 \\ 
	    1271 & 0 & 0 & 1 \\ 
	    1240 & 1 & 0 & 3\\ 
	    1220 & 1 & 0 & 5 \\
	    325 & 1 & 0 & 1 \\ 
         1262 &  0 & 0 & 3 \\
        \bottomrule
    \end{tabular}
    \caption{Comparing our success in writing assertions to find known bugs in
	the Hack@DAC2018 design to HardFails ~\cite{dessouky2019fails}. }

    \label{tab:reproduce-hardfails}.
\end{table}

\section{Conclusion}
In this paper, we address a critical gap in the hardware security community: the
lack of open-source properties for verification. 
We demonstrate the importance of properties for the reproducibility of evaluation results 
through a case study with the Hack@DAC 2018 design. Additionally, we provide a set of 
SVA properties for four commonly used CPU and SoC designs as part of an open-source hardware 
verification benchmarks repository. The benchmarks we develop
organize these properties by design, detailing any native or inserted 
security flaws in each, and documenting the associated CWEs. Finally, we
provide the methodology we use for manual property writing and address some of the associated
challenges to enable further collaboration and advancement in this space.

\bibliographystyle{ACM-Reference-Format}

\bibliography{references}
\clearpage
\end{document}